\documentclass[12pt]{amsart}
\usepackage{amssymb,multicol,amscd,amsmath}

\def\vstavka[#1]{{\bf #1}}

\usepackage{hyperref}

\makeatletter \@addtoreset{equation}{section} \makeatother

\newcommand{\Add}[3]{\setcounter{#1}{#2}\addtocounter{#1}{#3}}
\newcommand{\MAdd}[6]{\Add{#1}{#2}{#3}\addtocounter{#1}{#4}\addtocounter{#1}{#5}\addtocounter{#1}{#6}}

\newcommand{\MulFive}[2]{\MAdd{#1}{#2}{#2}{#2}{#2}{#2}}
\newcommand{\MulTen}[2]{\MulFive{#1}{#2}\addtocounter{#1}{\value{#1}}}

\newcounter{PictureWidth}     \newcounter{PictureHeight}  

\newcounter{YoungWidth}\newcounter{YoungHeight}  
\newcounter{ULYoungX}\newcounter{ULYoungY}  
\newcounter{URYoungX}\newcounter{URYoungY}  
\newcounter{DLYoungX}\newcounter{DLYoungY}  

\newcounter{TempYoungWidth}\newcounter{TempYoungHeight}  
\newcounter{TempURYoungX}\newcounter{TempDLYoungY}       

\newcounter{CurrentYoungWidth}\newcounter{CurrentYoungHeight}
\newcounter{CurrentURYoungX}\newcounter{CurrentURYoungY}
\newcounter{CurrentDLYoungX}\newcounter{CurrentDLYoungY}

\newcounter{Temp}\newcounter{TempA}\newcounter{TempB}

\newenvironment{Young}[3]{#3\InitializeYoung{#1}{#2}\SetToRow{1}\SetToCol{1}%
\begin{picture}(\value{PictureWidth},\value{PictureHeight})}{\end{picture}}

\newcommand{\InitializeYoung}[2]{{%
\MulTen{Temp}{#1}\MAdd{PictureWidth}{\LSkipValue}{\value{Temp}}{\RSkipValue}{0}{0}%
\MulTen{Temp}{#2}\MAdd{PictureHeight}{\USkipValue}{\value{Temp}}{\DSkipValue}{0}{0}%
\setcounter{YoungWidth}{0}\setcounter{YoungHeight}{0}%
\setcounter{ULYoungX}{\LSkipValue}\Add{ULYoungY}{\value{PictureHeight}}{-\USkipValue}%
\setcounter{URYoungX}{\value{ULYoungX}}\setcounter{URYoungY}{\value{ULYoungY}}%
\setcounter{DLYoungX}{\value{ULYoungX}}\setcounter{DLYoungY}{\value{ULYoungY}}%
\setcounter{CurrentYoungWidth}{0}\setcounter{CurrentYoungHeight}{0}%
\setcounter{CurrentURYoungX}{\value{ULYoungX}}\setcounter{CurrentURYoungY}{\value{ULYoungY}}%
\setcounter{CurrentDLYoungX}{\value{ULYoungX}}\setcounter{CurrentDLYoungY}{\value{ULYoungY}}%
\ZeroTempCounters}}

\newcommand{\ZeroTempCounters}{\setcounter{TempYoungWidth}{0}\setcounter{TempYoungHeight}{0}%
\setcounter{TempURYoungX}{0}\setcounter{TempDLYoungY}{0}}

\newcounter{HLines}\newcounter{VLines}

\newcommand{\LSkipValue}{0}\newcommand{\RSkipValue}{0}  
\newcommand{\USkipValue}{0}\newcommand{\DSkipValue}{0}  

\newcounter{MiddleX}\newcounter{MiddleY}\newcounter{TempMiddleA}\newcounter{TempMiddleB}

\newcounter{CurrentRow}\newcounter{CurrentCol}
\newcounter{CurrentRowY}\newcounter{CurrentColX}

\newcommand{\SetToRow}[1]{{\setcounter{CurrentRow}{#1}\MulTen{Temp}{#1}%
\Add{CurrentRowY}{\value{ULYoungY}}{-\value{Temp}}}}

\newcommand{\SetToCol}[1]{\setcounter{CurrentCol}{#1}\MulTen{Temp}{#1}%
\MAdd{CurrentColX}{\value{ULYoungX}}{\value{Temp}}{-10}{0}{0}}

\newcounter{LabelX}\newcounter{LabelY}
\newcounter{Index}

\newcounter{HLineLength}\newcounter{VLineLength}
\newcounter{HLineRef}\newcounter{VLineRef}
\newcounter{HLineDots}\newcounter{VLineDots}

\newsavebox{\HLineBox}\newsavebox{\VLineBox}
\newsavebox{\HatchBox}

\newcounter{DotsPerBox}

\newcounter{BraceX}\newcounter{BraceY}\newcounter{BraceSize}

\newcounter{MBlockHeight}   

\def\Distance{12}\def\HalfDistance{6}
\newcounter{LengthRow} \newcounter{XRow} \newcounter{YRow} \newcounter{XEnd} \newcounter{YEnd}
\newcounter{XDelta}  \addtocounter{XDelta}{\Distance}
\newsavebox{\Elem}\sbox{\Elem}{\put(0,0){\circle*{3}}}%
\newsavebox{\ElemArrows}\sbox{\ElemArrows}{\put(0,0){\vector(-1,1){\Distance}}\put(0,0){\vector(1,1){\Distance}}\put(0,0){\usebox{\Elem}}}%
\newsavebox{\ArrowsElem}\sbox{\ArrowsElem}{\put(-\Distance,-\Distance){\vector(1,1){\Distance}}\put(\Distance,-\Distance)%
{\vector(-1,1){\Distance}}\put(0,0){\usebox{\Elem}}}%
\newsavebox{\Etc}\sbox{\Etc}{\linethickness{0.8pt}\qbezier[3](0,0)(\HalfDistance,\HalfDistance)(\Distance,\Distance)}%
\newsavebox{\ArrowElem}\sbox{\ArrowElem}{\put(0,0){\usebox{\Elem}}\put(0,0){\vector(1,1){\Distance}}\put(\Distance,\Distance){\usebox{\Elem}}} %
\newsavebox{\ArrowElemArrow}\sbox{\ArrowElemArrow}{\put(0,0){\usebox{\ArrowElem}}\put(\Distance,\Distance){\vector(-1,1){\Distance}}}%
\newcommand{\half}{\frac{1}{2}}

\def\be{\begin{equation}\begin{aligned}}
\def\ee{\end{aligned}\end{equation}}

\def\bal{\begin{align}}
\def\eal{\end{align}}

\def\ptl{\partial}
\newcommand{\dpd}[1]{\frac{\ptl}{\ptl #1}}
\newcommand{\ddpd}[2]{\frac{\ptl^2}{\ptl #1 \ptl #2}}

\def\cD{\mathcal{D}}  
  
 \def\cK{\mathcal{K}} \def\cL{\mathcal{L}}
\def\cM{\mathcal{M}}  
\def\cP{\mathcal{P}}

 \def\cZ{\mathcal{Z}}

\newcommand{\ao}{\mathfrak{o}}

\newcommand{\au}{\mathfrak{u}}
\newcommand{\asp}{\mathfrak{sp}}

\newcommand{\aiu}{\mathfrak{iu}}
\newcommand{\aisu}{\mathfrak{isu}}

\newcommand{\ahsc}{\mathfrak{hsc}(4)}

\newcommand{\acu}{\mathfrak{cu}(1,0|8)}
\newcommand{\ahuzero}{\mathfrak{hu}_0(1,0|8)}

\def\ideal{\mathfrak{I}}

\def\uL{\cL} \def\uD{\cD} \def\uP{\cP} \def\uK{\cK} 
\def\buL{\bar{\uL}}

\def\adinfL{\uL{}} \def\adinfbL{\buL{}}
\def\adinfP{\uP{}} \def\adinfK{\uK{}} \def\adinfD{\uD{}} \def\adinfPNO{\PNO{}}

\def\adP{\uP{}} \def\adK{\uK{}}

\def\trProj{\Pi^\bot}
\def\ttrProj{\tilde{\Pi}^\bot}

\def\na{n_a} \def\nb{n_b} \def\nba{n_\ba} \def\nbb{n_\bb}
\def\bn{\bar n}  \def\tbn{\bar{\tilde{n}}}

\def\tb{\tilde{b}} 
\def\tbb{\bar{\tilde{b}}} \def\ntbb{n_{\tbb}}

\def\twinfL{\tilde{\uL}} \def\twinfbL{\tilde{\buL}}
\def\twinfP{\tilde{\uP}} \def\twinfK{\tilde{\uK}} \def\twinfD{\tilde{\uD}} \def\twinfPNO{\tilde{\PNO}}

\def\twP{\twinfP}

\def\btwP{\bar{\twP}}

\def\tD{\tilde{\uD}}

\def\adbasisanscoef{d}

\def\twbasisanscoef{\tilde{d}}

\def\adbasefunc{g}
\def\twbasefunc{\tilde{g}}

\def\admodule{\cM}
\def\admodules[#1]{\admodule_{#1}}
\def\admoduleinf{\admodule^\infty}
\def\admoduleinfs[#1]{\admoduleinf_{#1}}

\def\twmodule{\tilde{\cM}}
\def\twmodules[#1]{\twmodule_{#1}}
\def\twmodulen[#1]{\twmodule^{#1}}
\def\twmodulens[#1][#2]{\twmodule^{#1}_{#2}}
\def\twmoduleinf{\twmodule^\infty}
\def\twmoduleinfs[#1]{\twmoduleinf_{#1}}

\def\btwmodule{\bar{\tilde{\cM}}}
\def\btwmodules[#1]{\btwmodule_{#1}}
\def\btwmodulen[#1]{\btwmodule^{#1}}
\def\btwmodulens[#1][#2]{\btwmodule^{#1}_{#2}}
\def\btwmoduleinf{\btwmodule^\infty}
\def\btwmoduleinfs[#1]{\btwmoduleinf_{#1}}

\def\twsubmodulen[#1]{\tilde{\ideal}^{#1}}
\def\twsubmodulens[#1][#2]{\tilde{\ideal}^{#1}_{#2}}

\def\opsinf{{s}}

\def\gaugepar{\varepsilon}
\def\physf{\varphi}

\def\Weyltens{C}
\def\tweq{\tilde{E}}

\def\extdiff{{\rm d}}

\def\ga{\alpha}
\def\gb{\beta}
\def\gga{\gamma}
\def\gd{\delta}
\def\gs{\sigma}

\def\gep{\epsilon}

\def\da{{\dot \ga}}
\def\db{{\dot \gb}}
\def\dg{{\dot \gga}}
\def\dd{{\dot \gd}}

\def\ba{{\bar a}}
\def\bb{{\bar b}}

\def\rewcompmap[#1]{\left.#1\right|_{\stackrel{\psi=\xi}{c=b}}}

\def\Ch[#1]{\compel^{p,(#1)}_{s_1,s_2,N}}

\def\eh[#1]{\epsilon^{p-1,(#1)}_{s_1,s_2,N}}
\def\tCh[#1]{\tilde{\compel}^{p,(#1)}_{s_1,s_2,N}}

\def\bt{\begin{tabular}}
\def\et{\end{tabular}}

\newcommand{\compel}{C}

\def\PNO{\cZ}

\def\Curv{R}
\def\GaugeF{\omega}
\def\WeylF{C}
\def\bWeylF{\bar{C}}
\def\adsm{\gs_-}
\def\twsm{\tilde{\gs}_-}
\def\btwsm{\bar{\tilde{\gs}}_-}
\def\ChEsm{\gs{}}
\def\bChEsm{\bar{\gs}{}}

\begin{document}

\title{Bosonic Fradkin-Tseytlin equations unfolded. Irreducible case.}

\author{O.V.~Shaynkman}

\maketitle
\vspace{-20pt}
\begin{center}
{\small\it I.E.Tamm Theory Department, Lebedev Physical Institute, Leninski
prospect 53,\\ 119991, Moscow, Russia}
\end{center}
\begin{abstract}
We factorize 4d Fradkin-Linetsky higher spin conformal algebra by maximal ideal $\ideal^1-\ga$ and construct irreducible infinite-dimensional modules
$\admodule_\ga$ of 4d conformal algebra that are parameterized by real number $\ga$.
It is shown that independently of $\ga$ unfolded system of equations corresponding to each $\admodule_\ga$ describes collection of Fradkin-Tseytlin equations
for all spins $s=1,\dots,\infty$ with zero multiplicity.
\end{abstract}

\footnotetext{\scriptsize{\tt e-mail: shayn@lpi.ru}}

\section{Introduction\label{Introduction}}

Conformal higher spin theory, i.e.  theory of interacting conformal fields of all spins, is an interesting object from both AdS/CFT correspondence and  
possible connection to AdS higher spin theory points of view. It was discussed in a quite amount of papers 
(some of which are \cite{Pope}-\cite{Bekaert_Grigoriev})
starting from the pioneering work by Fradkin and Tseytlin \cite{Fradkin_Tseytlin}, where Lagrangians and corresponding equations of motion 
(Fradkin-Tseytlin equations) describing 
free dynamics of conformal fields of all spins where constructed. Then in paper \cite{Fr_Lin_conf_HS_alg} 
infinite dimensional algebras were introduced that were argued 
to be a proper candidates for the role of conformal higher spin algebra. 

The construction of \cite{Fr_Lin_conf_HS_alg} was as follows.
Consider Weyl star product algebra of two-component oscillators 
$a, b, \ba, \bb$ 
\begin{equation}\label{oscillators}
    [b_\gb,a^\ga]_*=\gd^\ga_\gb\,, \quad [\bb_\db,\ba^\da]_*=\gd^\da_\db\,,\quad \ga,\da=1,2\,,
\end{equation}
with the standard star product defined for symbols of operators $f(a,b,\ba,\bb)$ and $g(a,b,\ba,\bb)$ by formula
$f*g=f\exp(\overleftrightarrow{\Delta})g\,,$ where
\begin{equation}\label{star_product_Delta}
  \overleftrightarrow{\Delta}=\half\left(\overleftarrow{\dpd{b}}\cdot\overrightarrow{\dpd{a}}-
                                         \overleftarrow{\dpd{a}}\cdot\overrightarrow{\dpd{b}}+
                                         \overleftarrow{\dpd{\bb}}\cdot\overrightarrow{\dpd{\ba}}-
                                         \overleftarrow{\dpd{\ba}}\cdot\overrightarrow{\dpd{\bb}}\right)\,.
\end{equation}
As was shown in \cite{Gunaydin}, \cite{Bars_Gunaydin},  bilinear combinations of oscillators with respect to star product commutator $[f,g]_*$ 
form $\asp(8)$ algebra, which reduces to 
4d conformal subalgebra $\au(2,2)$ when restricted by two additional conditions 
\begin{equation}\label{iucond}
\begin{aligned}
&\mbox{1.}&& \mbox{Centralization by helicity operator }\PNO=i/2(a^\ga b_\ga-\ba^\da \bb_\da);\quad [f,\PNO]_*=0\\
&\mbox{2.}&& \mbox{Reality condition }\quad f(a,b,\ba,\bb)=-\bar{f}(i\ba,i\bb,ia,ib).
\end{aligned}
\end{equation}
The idea of Fradkin and Linetsky was to bring all polynomials (not only bilinear) into
the play but still keep conditions \eqref{iucond} imposed and, thus, get infinite-dimensional extension of $\au(2,2)$, which they called $\aiu(2,2)$. 

Let us note that algebra $\aiu(2,2)$ is isomorphic to $AdS_5$ higher spin algebra that was discussed in several papers
\cite{Vasiliev_AdS_5_cubic}-%
\cite{Alkalaev_FV_type_AdS_5}. In \cite{Vasiliev_conf_HS_sym in _four_dimm} it was denoted as
$\acu$ where 8 indicates the number of oscillators used and pair 1,0 points out that it has trivial structure in spin 1 Yang-Mills sector.
Algebra $\aisu(2,2)$ (i.e. $\aiu(2,2)$ factorized by all star powers of $\PNO$)  was originally (in \cite{Fr_Lin_conf_HS_alg}) denoted as $\ahsc$,
where hsc means higher spin conformal and 4 indicates that it extends 4-dimensional conformal algebra. It is isomorphic to the
minimal $AdS_5$ higher spin algebra denoted as $\ahuzero$ in \cite{Vasiliev_conf_HS_sym in _four_dimm}.
As was discussed in \cite{Eastwood_Symmetries_of_Laplasian}-
\cite{Joung_Mkrtchan_Notes_on_HS_algebras}
one can associate the minimal $AdS_5$ higher spin algebra with the quotient of universal enveloping $AdS_5$ Lie algebra
over the kernel of its singleton representation.

By analogy with the $AdS$ case the procedure of construction of full nonlinear conformal higher spin theory could be separated into two steps:
\begin{enumerate}
\item Reformulate linear equations of motion in unfolded (first order) form with higher spin algebra been gauge symmetry algebra of the system.
\item Deform nonlinearly both equations of motion and gauge symmetries in the self consistent way.   
\end{enumerate}
Although the second step requires a big amount of guess, the first step is rather straightforward. In paper \cite{Vasiliev_progr_in_HS_gauge_theory}
unfolded formulation of Fradkin-Tseytlin equations corresponding to reducible algebra $\aiu(2,2)$ was given and in paper \cite{OVS_Fr_Ts_Unf}
the spectrum of spins described by this system was obtained. At the present paper we contract system of \cite{Vasiliev_progr_in_HS_gauge_theory} to irreducible 
case and analyse it spectrum of spins. 

More precisely, as we already mentioned, bilinear combinations of oscillators that are coordinated with \eqref{iucond} 
form 4 dimensional conformal algebra $\ao(4,2)\sim\au(2,2)\subset\aiu(2,2)$.
In paper \cite{OVS_Fr_Ts_Unf} adjoint and twisted-adjoint%
\footnote{Twisted-adjoint module can be thought as Fourier transform of adjoint module (see section \ref{Tw_Mod}) and \cite{OVS_Fr_Ts_Unf} fore more details.} 
modules $\admodule^\infty$ and
$\twmodule^\infty$ of $\au(2,2)$ acting on $\aiu(2,2)$ were constructed and analyzed in rather details.
It was show that both modules decompose into direct sum of infinitely many copies of irreducible submodules $\admodule_s$, $\twmodule_s$
corresponding to spins $s=1,2,\ldots$
\begin{equation}\label{decompos}
\begin{aligned}
&\admodule^\infty=\oplus_{s=1}^\infty\infty\admodules[s]\,,\\
&\twmodule^\infty=\oplus_{s=1}^\infty\infty\twmodules[s]\,.
\end{aligned}
\end{equation}
Here module $\admodule^\infty$ ($\twmodule^\infty$) contain submodules $\admodule_s$ ($\twmodule_s$) in infinitely many copies, which enter
$\admodule^\infty$ ($\twmodule^\infty$) with the factor $(\PNO*)^n$ ($(\tilde{\PNO}*)^n$),
$n=0,1,\ldots \infty$ (recall that $\aiu(2,2)$ is a centralizer of $\PNO$).

It was shown that unfolded system constructed with respect to modules $\admodule^\infty$,
$\twmodule^\infty$ is coordinated with decomposition \eqref{decompos} and, thus, decomposes into infinitely many copies of unfolded systems corresponding to
free conformal equations on spin $s$ field $s=1,2,\ldots$ (Fradkin-Tseytlin equations).

The degeneration of unfolded system considered in \cite{OVS_Fr_Ts_Unf} is due to non simplicity of algebra $\aiu(2,2)$.
Really, it contains an infinite chain of ideals
\begin{equation}\label{isukk_ideals_embedded}
  \aiu(2,2)\supset\ideal^1\supset\ideal^2\supset\cdots\supset\ideal^m\supset\cdots\,,
\end{equation}
generated by star powers of $\PNO$, i.e. ideal $\ideal^m$ is spanned by the elements of form $(\PNO*)^m*h(a,b,\ba,\bb)$. In \cite{OVS_Fr_Ts_Unf} it was speculated
that unfolded system constructed with respect to irreducible algebra $\aisu_0(2,2)=\aiu(2,2)/\ideal^1$ should contain each spin $s=1,2,\ldots$ in one copy only.

In the present paper we consider this case in more general formulation. Namely one can construct a series of ideals $\ideal^1_\ga$ of algebra
$\aiu(2,2)$ that are spanned by the elements
of form $(\PNO-\ga)*h(a,b,\ba,\bb)$ for some real number $\ga$. Quotients $\aiu(2,2)/\ideal^1_\ga$ give rise to the series of irreducible
infinite-dimensional algebras $\aisu_\ga(2,2)$. We briefly discuss the structure
of adjoint and twisted-adjoint $\au(2,2)$-modules on $\aisu_\ga(2,2)$. Consider
unfolded system of equations corresponding to these modules and show that, as was speculated in \cite{OVS_Fr_Ts_Unf}
it is decomposed into the subsystems corresponding to Fradkin-Tseytlin equations
for one copy of every spin $s=1,2,\ldots$ independently of $\ga$.

The rest of the paper is rater technical and we refer to \cite{OVS_Fr_Ts_Unf}
for more details.

\section{Adjoint module}
Adjoint action of $\au(2,2)$-generators on $\aiu(2,2)$ is given by formulas
\begin{equation}\label{adinf_repr_ukk}
  \begin{aligned}
    &\adinfL_\ga{}^\gb=a^\gb\dpd{a^\ga}-b_\ga\dpd{b_\gb}-\frac{1}{2}\gd_\ga^\gb(\na-\nb)\,,\\
    &\adinfbL_\da{}^\db=\ba^\db\dpd{\ba^\da}-\bb_\da\dpd{\bb_\db}-\frac{1}{2}\gd_\da^\db(\nba-\nbb)\,,\\
    &\adinfP_{\ga\db}=b_\ga\dpd{\ba^\db}+\bb_\db\dpd{a^\ga}\,,\quad
    \adinfK^{\ga\db}=-a^\ga\dpd{\bb_\db}-\ba^\db\dpd{b_\ga}\,,\\
    &\adinfD=\frac{1}{2}(\na+\nba-\nb-\nbb)\,,\quad
    \adinfPNO=\frac{i}{2}(\na-\nba-\nb+\nbb)\,,
  \end{aligned}
\end{equation}
where $n_a\,,n_b\,,n_\ba\,,n_\bb$ are Euler operators for corresponding oscillators. Generators \eqref{adinf_repr_ukk} commute with
spin operator
\be\label{spinop}
\opsinf=\na+\nbb+1=\nb+\nba+1\,,
\ee
which, thus, decomposes the whole $\au(2,2)$-module into the submodules with the spin $s$ fixed.

To find adjoint representation of $\au(2,2)$ on quotient algebra $\aisu_\ga(2,2)$ one should fix some basis on $\aisu_\ga(2,2)$ first.
The most natural basis is
\be\label{iuga_basis_nat}
\uD^v f(a,b,\ba,\bb)\,,\quad v=0,1,\ldots
\ee
where $\uD=1/2(a^\ga b_\ga+\ba_\da\bb_\da)$ is operator of dilatation and $f(a,b,\ba,\bb)$ is polynomial of oscillators, which
satisfy conditions \eqref{iucond} and is traseless, i.e. satisfy relations
\begin{equation}\label{trcond}
  \ddpd{a^\ga}{b_\ga}f=0\,,\quad \ddpd{\ba^\da}{\bb_\da}f=0.
\end{equation}
However, generators of $\au(2,2)$, namely generators of translation and special conformal transformation, are not diagonal with respect to
spin value in this basis. Really factorization requirement $(\PNO-\ga)*h(a\,,b\,,\ba\,,\bb)\sim 0$ doesn't commute with spin operator \eqref{spinop}.

To avoid this inconvenience let us consider more general ansatz for the basis elements
\be\label{iuga_basis}
g_s^v(\uD) f(a,b,\ba,\bb)\,,\quad s=1,2\ldots\,,\quad v=0,1,\ldots, s-1\,,
\ee
where
\be
\label{gfunc}
g^v_s(\uD)=\sum_{j=0}^v\uD^j\adbasisanscoef^v_{s;j}\,,\qquad \adbasisanscoef^v_{s;v}\equiv 1
\ee
is some polynomial on $\uD$ of the power $v$ with coefficients $\adbasisanscoef^v_{s;j}$ to be found.

Direct computation (analogues to that in \cite{OVS_Fr_Ts_Unf}) brings us to the following result
\be
\label{gfunccoef}
\adbasisanscoef^v_{s;j}={v\choose j}\frac{\gd_{s;v-j}}{\prod_{h=1}^{v-j}(2s-h+1)}\,,
\ee
where $\gd_{s;m}$ are some functions depending on $i(n-\bn)$ with $n=n_a+n_b$, $\bn=n_\ba+n_\bb$.
Functions $\gd_{s;m}$ are fixed by the following recurrence equation
\be
\label{gfunccoefgd}
\gd_{s;m}=i\ga(n-\bn)\gd_{s;m-1}+\frac{(m-1)}{8}(2s-m+2)(s-v-8\ga^2+1)\gd_{s;m-2}\,,\quad m=1,\ldots, v\,,
\ee
with boundary conditions $\gd_{s;0}\equiv 1\,, \quad \gd_{s;m<0}\equiv 0$.

In new basis \eqref{iuga_basis} operators $\adinfL_\ga{}^\gb$, $\adinfbL_\da{}^\db$, $\adinfD$, $\adinfPNO$ are given by the same formulae \eqref{adinf_repr_ukk} 
while operators $\adinfP_{\ga\db}$ and $\adinfK^{\ga\db}$ have the following new form
\begin{equation}\label{ad_repr_ukk}
  \begin{aligned}
    &\adP_{\ga\db}\:\adbasefunc^v_{s}f=\left[v\adbasefunc^{v-1}_{s}\trProj b_\ga\bb_\db+
\frac{v+n+2}{n+2}\adbasefunc^v_{s}\trProj\bb_\db\dpd{a^\ga}+\frac{v+\bn+2}{\bn+2}\adbasefunc^v_{s}\trProj b_\ga\dpd{\ba^\db}+\right.\\
    &\quad\qquad\qquad\qquad\qquad\qquad\qquad\qquad\qquad\qquad\qquad\left.{}+\frac{2s-v}{(n+2)(\bn+2)}\adbasefunc^{v+1}_{s}
\ddpd{a^\ga}{\ba^\db}\right]f\,,\\
    &\adK^{\ga\db}\:\adbasefunc^v_{s}f=-\left[v\adbasefunc^{v-1}_{s}\trProj a^\ga\ba^\db+
\frac{v+n+2}{n+2}\adbasefunc^{v}_{s}\trProj\ba^\db\dpd{b_\ga}+\frac{v+\bn+2}{\bn+2}\adbasefunc^{v}_{s}\trProj a^\ga\dpd{\bb_\db}+\right.\\
    &\qquad\qquad\qquad\qquad\qquad\qquad\qquad\qquad\qquad\qquad\qquad\left.+\frac{2s-v}{(n+2)(\bn+2)}\adbasefunc^{v+1}_{s}
\ddpd{b_\ga}{\bb_\db}\right]f\,,
  \end{aligned}
\end{equation}
where $\trProj=(\trProj)^2$ is projector to the traceless components \eqref{trcond}.

Operators \eqref{ad_repr_ukk} commute with spin operator $s$, which in new basis has form
\be\label{newspinop}
\opsinf=v+\na+\nbb+1=v+\nb+\nba+1\,.
\ee
We, thus, have that all modules $\admodule_\ga$ of $\au(2,2)$ adjoint action on algebra $\aisu_\ga(2,2)$ are isomorphic and decompose into
direct sum
\begin{equation}\label{addecompos}
\admodule_\ga=\oplus_{s=1}^\infty\admodules[s]\,,
\end{equation}
where $\admodules[s]$ are finite-dimensional $\au(2,2)$-submodules of $\admodule_\ga$, which are spanned by basic vectors \eqref{iuga_basis} with
$s$ fixed.

\section{Twisted-adjoint module\label{Tw_Mod}}
Twisted-adjoint module $\twmodule_\ga$ can be obtained from adjoint module $\admodule_\ga$ by twist transformation
\be\label{twist}
\bb_\da&\rightarrow\dpd{\tbb^\da}\,,\\
\dpd{\bb_\da}&\rightarrow-\tbb^\da\,,
\ee
which preserves commutator
\be\label{twist_commut_conserv}
[\dpd{\bb_\da},\bb_\db]=[-\tbb^\da,\dpd{\tbb^\db}]\,.
\ee

Under transformation \eqref{twist} elements \eqref{iuga_basis} get the following form
\be\label{iuga_twbasis}
\tilde{g}_s^v(\tD) \tilde{f}(a,b,\ba,\tbb)\,,\quad s=1,2\ldots\,,\quad v=0,1,\ldots, s-1\,,
\ee
where $\tilde{f}$ satisfies twisted-traceless conditions
\be\label{twtrcond}
\ddpd{a^\ga}{b_\ga}f=0\,,\quad \tbb^\da\dpd{\ba^\da}f=0
\ee
and $\tD=1/2(a^\ga b_\ga+\ba^\da\dpd{\tbb^\da})$. Here $\tilde{g}_s^v(\tD)$ is a polynomial of $\tD$
\be
\label{tgfunc}
\tilde{g}^v_s(\tD)=\sum_{j=0}^v\tD^j\twbasisanscoef^v_{s;j}\,,\qquad \twbasisanscoef^v_{s;v}\equiv 1\,,
\ee
coefficients $\twbasisanscoef^v_{s;j}$ are given by
\be
\label{tgfunccoef}
\twbasisanscoef^v_{s;j}={v\choose j}\frac{\tilde{\gd}_{s;v-j}}{\prod_{h=1}^{v-j}(2s-h+1)}\,,
\ee
and $\tilde{\gd}_{s;m}$ are some functions depending on $i(n-\tbn)$, where $n=n_a+n_b$ is the same as in the previous section and $\tbn=n_\ba-n_{\tbb}-2$ is
twist-transformed operator $\bn$. Functions $\tilde{\gd}_{s;m}$ are fixed by the following recurrence equation
\be
\label{tgfunccoefgd}
\tilde{\gd}_{s;m}=i\ga(n-\tbn)\tilde{\gd}_{s;m-1}+\frac{(m-1)}{8}(2s-m+2)(s-v-8\ga^2+1)\tilde{\gd}_{s;m-2}\,,\quad m=1,\ldots, v\,,
\ee
with boundary conditions $\tilde{\gd}_{s;0}\equiv 1\,, \quad \tilde{\gd}_{s;m<0}\equiv 0$.

The structure of module $\twmodule_\ga$ differs from that of $\admodule_\ga$.
Firstly, elements \eqref{iuga_twbasis} are not linearly independent. Really, due to \eqref{twtrcond} function $\tilde{f}$ forms two-row
Young tableau with respect to dotted indices with fist (second) row of length $n_{\tbb}$ ($n_\ba$). Thus, operator $(\tbb^\da\dpd{\ba^\da})^u$ vanishes
on $\tilde{f}$ for $u>v_{\max}=n_{\tbb}-n_\ba$. Therefore, collection of linearly independent elements of \eqref{iuga_twbasis},
which form basis in $\twmodule_\ga$ is given by formula \eqref{iuga_twbasis} but with $v$ bounded by $\min(s-1,v_{\max})$ (cf. \eqref{iuga_basis})
\be\label{iuga_twbasis_real}
\tilde{g}_s^v(\tD) \tilde{f}(a,b,\ba,\tbb)\,,\quad s=1,2\ldots\,,\quad v=0,1,\ldots, \min(s-1,v_{\max})\,.
\ee

Secondly, since the change of the sign in front of $n_{\tbb}$, twisted-spin operators have form
\be\label{twspin}
\tilde{s}=n_a-n_{\tbb}+v-1=n_b+n_\ba+v+1
\ee
with subtraction of $n_a$ and $n_{\tbb}$. Thus, fixation of the spin value $s$ doesn't bound the order of $\tilde{f}$ with respect to oscillators $a$ and
$\tbb$. This means that $\au(2,2)$-modules $\twmodule_s$, which are formed by \eqref{iuga_twbasis_real} with fixed $s$, are infinite-dimensional contrary to $\admodule_s$.

Applying twisted transformation \eqref{twist} to formulae \eqref{adinf_repr_ukk}, \eqref{ad_repr_ukk} one finds $\au(2,2)$-representation on $\twmodule_s$
\begin{equation}\label{tw_repr}
  \begin{aligned}
    &\twinfL_\ga{}^\gb=a^\gb\dpd{a^\ga}-b_\ga\dpd{b_\gb}-\frac{1}{2}\gd_\ga^\gb(\na-\nb)\,,\\
    &\twinfbL_\da{}^\db=\ba^\db\dpd{\ba^\da}+\tbb^\db\dpd{\tbb^\da}-\frac{1}{2}\gd_\da^\db(\nba+\ntbb)\,,\\
    &\twinfP_{\ga\db}\:\twbasefunc^v_{s}\tilde{f}=\left[v\twbasefunc^{v-1}_{s}\ttrProj b_\ga\dpd{\tbb^\db}+
\frac{v+n+2}{n+2}\twbasefunc^v_{s}\ttrProj\ddpd{a^\ga}{\tbb^\db}+\frac{v+\tbn+2}{\tbn+2}\twbasefunc^v_{s}\ttrProj b_\ga\dpd{\ba^\db}+\right.\\
    &\quad\qquad\qquad\qquad\qquad\qquad\qquad\qquad\qquad\qquad\qquad\left.{}+\frac{2s-v}{(n+2)(\tbn+2)}\twbasefunc^{v+1}_{s}
\ddpd{a^\ga}{\ba^\db}\right]\tilde{f}\,,\\
    &\twinfK^{\ga\db}\:\twbasefunc^v_{s}\tilde{f}=-\left[v\twbasefunc^{v-1}_{s}\ttrProj a^\ga\ba^\db+
\frac{v+n+2}{n+2}\twbasefunc^{v}_{s}\ttrProj\ba^\db\dpd{b_\ga}-\frac{v+\tbn+2}{\tbn+2}\twbasefunc^{v}_{s}\ttrProj a^\ga\tbb^\db+\right.\\
    &\qquad\qquad\qquad\qquad\qquad\qquad\qquad\qquad\qquad\qquad\qquad\left.+\frac{2s-v}{(n+2)(\tbn+2)}\twbasefunc^{v+1}_{s}
\tbb^\db\dpd{b_\ga}\right]\tilde{f}\,,\\
    &\twinfD=\frac{1}{2}(\na+\nba-\nb+\ntbb+2)\,,\quad
    \twinfPNO=\frac{i}{2}(\na-\nba-\nb-\ntbb-2)\,,
  \end{aligned}
\end{equation}
where $\ttrProj=(\ttrProj)^2$ is projector to component satisfying \eqref{twtrcond}.

It is almost obvious that operators \eqref{tw_repr} commute with spin operator $\tilde{s}$ and, thus, module
$\twmodule_\ga$ admit decomposition analogous to that of adjoint case
\begin{equation}\label{twdecompos}
\twmodule_\ga=\oplus_{s=1}^\infty\twmodules[s]\,.
\end{equation}
Though it is worth to mention that for operators of translation and of special conformal transformation some additional
analyses is needed. Really, the second terms of these operators, when acting on the element $\tilde{g}^{v_{\max}}_s\tilde{f}$ with the maximal value of $v$,
decrease the value of $v_{\max}$ by one but still don't decrease $v$ and, thus, break the limit $v\leq v_{\max}$. However, it can be shown that
coefficients in front of these problematic terms zero out and therefore decomposition \eqref{twdecompos} holds (see \cite{OVS_Fr_Ts_Unf} for details).

Finally, let us note that we also need complex conjugated twisted-adjoint module $\bar{\twmodule}_\ga=-\twmodule_\ga^\dag$,
where $\dag$ is Hermitian involution defined by the following relations on oscillators
\be\label{inv}
& (a^\ga)^\dag=i\ba^\da\,,\quad (b_\ga)^\dag=i\bb_\da\,,\quad (\ba^\da)^\dag=ia^\ga\,,\quad (\tbb^\da)^\dag=i\tb^\ga\,,\\
& f^\dag(a,b,\ba,\tbb)=\bar{f}(a^\dag,b^\dag,\ba^\dag,\tbb^\dag)\,.
\ee

\section{Unfolded system}
Unfolded system for the family of Fradkin-Linetsky higher spin conformal algebras was first formulated for algebra $\aiu(2,2)$ in paper
\cite{Vasiliev_progr_in_HS_gauge_theory}. Factorizing this system by ideal $\ideal_1-\ga$ one gets
\be\label{unfsys}
&\Curv=(\extdiff +\adsm)\GaugeF\,,\\
&\Curv=\ChEsm\WeylF+\bChEsm\bWeylF\,,\\
&(\extdiff+\twsm)\WeylF=0\,,\\
&(\extdiff+\btwsm)\bWeylF=0\,.
\ee
Let us briefly discuss the structure of this system.
\begin{itemize}
\item[1 eq.:] $\Curv$ is 2-form linearized higher spin curvature constructed from the set of 1-form gauge fields $\GaugeF$ taking values in
$\admodule_\ga$;
\be
\extdiff={\rm d}x^n\dpd{x^n}\,, n=0,\ldots, 4
\ee
is exterior differential; and
\be
\adsm=\xi^{\ga\db}\adP_{\ga\db}\,,
\ee
is adjoint part of sigma-minus operator (see. \cite{Vasiliev_unfolded_repr_for_eq_in_two_plus_one_AdS}-%
\cite{Shaynkman_Vasiliev_Scalar_field_from_the_HST_perspective}
for more details on unfolded formulation and $\gs_-$ technic) where
flat connection 1-form $\xi^{\ga\db}={\rm d}x^n\gs_n^{\ga\db}$ is constructed from 4 Pauli matrices.
\item[3,4 eq.:] $\WeylF$ and $\bWeylF$ are sets of 0-form Weyl tensors taking values in $\twmodule_\ga$ and $\btwmodule_\ga$ correspondingly; and
\be
\twsm=\xi^{\ga\db}\twP_{\ga\db}\,,&& \btwsm=\xi^{\ga\db}\btwP_{\ga\db}
\ee
are twisted-adjoint parts of sigma-minus operator.
\item[2 eq.:] Expresses nonzero parts of linearized curvature $\Curv$ through the Weyl tensors $\WeylF$ and $\bWeylF$. Here operator
$\ChEsm$ ($\bChEsm$), which maps 0-form $\WeylF$ ($\bWeylF$) into 2-form $\Curv$, is given by formula
\be
\ChEsm\tilde{g}_s^v\tilde{f}=(1/2)^v\xi^{\ga\db}\xi^{\ga\dg}\gep_{\db\dg}\ddpd{a^\ga}{a^\ga}\Big(\ba^\da\dpd{\tbb^\da} \Big)^v \tilde{f}\Big|_{\tbb=0}\,,\\
\bChEsm\tilde{g}_s^v\tilde{f}=(1/2)^v\xi^{\gb\da}\xi^{\gga\da}\gep_{\gb\gga}\ddpd{\ba^\da}{\ba^\da}\Big(a^\ga\dpd{\tb^\ga} \Big)^v \tilde{f}\Big|_{\tb=0}\,.
\ee
\end{itemize}

System \eqref{unfsys} split into subsystems with fixed spin $s=1,2,\ldots$. This is obviously true for
equations 1, 3 and 4 due to decompositions \eqref{addecompos} and \eqref{twdecompos} for modules $\admodule_\ga$, $\twmodule_\ga$ and $\btwmodule_\ga$.
To check this splitting for
the second equation one should show that operator $\ChEsm+\bChEsm$ maps 0-forms taking values in $\twmodule_s\oplus\btwmodule_s$ into 2-forms taking values in
$\admodule_s$. Really, let us consider operator $\ChEsm$, it doesn't vanish for fields $\tilde{g}^v_s\tilde{f}$, which are independent of $\ba$ and contain
$v$ oscillators $\tbb$ only. As one can easily see these are fields with
\be\label{ChEsmnonzerons}
n_a=s+1\,,\quad n_b=s-v-1\,,\quad n_\ba=0\,,\quad n_{\tbb}=v\,.
\ee
Operator $\ChEsm$ transforms \eqref{ChEsmnonzerons} into
\be\label{ChEsmmapstons}
n_a=s-1\,,\quad n_b=s-v-1\,,\quad n_\ba=v\,,\quad n_\bb=0\,.
\ee
For operator $\bChEsm$ one gets complex conjugated analogs of \eqref{ChEsmnonzerons} and \eqref{ChEsmmapstons}.
Substituting these values in \eqref{newspinop} one finds that $\ChEsm+\bChEsm$ maps 0-form under consideration into 2-for taking values in
$g^0_s f$. So it preserves the spin value and, thus, system \eqref{unfsys} really split into subsystems with fixed spin.

To analyze dynamical content of system \eqref{unfsys} one should note first that almost all equations of general unfolded system just
express higher order fields as derivatives of lower order fields. (In our case the order of fields is given by their conformal weight.)
And dynamically nontrivial aspects are hidden in cohomology of the operator $\hat{\gs}_-$ (see \cite{Shaynkman_Vasiliev_Scalar_field_from_the_HST_perspective}
for more details) $\hat{H}^p_{\hat{\gs}_-}$.

In our case
\be\label{hatsigmaminus}
\hat{\gs}_-=\adsm+\twsm+\btwsm+\ChEsm+\bChEsm
\ee
and
\begin{itemize}
  \item differential gauge parameters are given by $\hat{H}^0_{\hat{\gs}_-}$;
  \item dynamical fields are given by $\hat{H}^1_{\hat{\gs}_-}$;
  \item differential equations on dynamical fields are in one-to-one correspondence with
        $\hat{H}^2_{\hat{\gs}_-}$.
\end{itemize}

Note that, since $\hat{\gs}_-$ mixes $p$ and $p-1$-forms, $p$-th cohomology of $\hat{\gs}_-$ is a pair
$\hat{H}^p=(H^p,H^{p-1})$, which were found in \cite{OVS_Fr_Ts_Unf}.
Here we give a result:
\be\label{smcohom}
&\hat{H}^0_{\hat{\gs}_-}=(i^{s}\gaugepar^{\gb(s-1)\,;\,\db(s-1)}\,b_{\gb(s-1)}\bb_{\db(s-1)},0)\,,\\
&\hat{H}^1_{\hat{\gs}_-}=(i^{s}\xi^{\gga\dd}\,\physf^{\gb(s-1)\,;\,\db(s-1)}_{\gga\,;\,\dd}\,b_{\gb(s-1)}\bb_{\db(s-1)},0)\,,\\
&\hat{H}^2_{\hat{\gs}_-}=(0,i^{s}\xi^{\gga\dd}\,\twbasefunc^{s-1}_{s}\,\tweq_{\ga(s+1),\gga\,;\,\db(s-1)\dd}\,a^{\ga(s+1)}\tbb^{\db(s-1)}+c.c.)\,,
\ee
where $\gaugepar^{\gb(s-1)\,;\,\db(s-1)}$, $\physf^{\gb(s-1)\,;\,\db(s-1)}_{\gga\,;\,\dd}$
are gauge parameter, dynamical field for the spin $s$ subsystem of \eqref{unfsys} and
$\tweq_{\ga(s+1),\gga\,;\,\db(s-1)\dd}\,a^{\ga(s+1)}\tbb^{\db(s-1)}+c.c.=0$ corresponds to differential equations imposed on dynamical field.
These equations have the following form
\be
\label{spin_s_diff_eq}
&\Weyltens^{\ga(2s)}=\underbrace{\dpd{x_{\ga}{}^{\da}}\cdots\dpd{x_{\ga}{}^{\da}}}_{s}\physf^{\ga(s)\,;\,\da(s)}\,,
&&\bar{\Weyltens}^{\da(2s)}=\underbrace{\dpd{x^{\ga}{}_{\da}}\cdots\dpd{x^{\ga}{}_{\da}}}_{s}\physf^{\ga(s)\,;\,\da(s)}\,,\\
&\underbrace{\dpd{x^{\ga}{}_{\da}}\cdots\dpd{x^{\ga}{}_{\da}}}_{s}\Weyltens^{\ga(2s)}=0\,,
&&\underbrace{\dpd{x_{\ga}{}^{\da}}\cdots\dpd{x_{\ga}{}^{\da}}}_{s}\bar{\Weyltens}^{\da(2s)}=0\,,
\ee
where $\physf$ obeys the gauge transformations
\be
\label{spin_s_gauge_trans}
\gd\physf^{\ga(s)\,;\,\db(s)}=\dpd{x_{\ga\db}}\gaugepar^{\ga(s-1)\,;\,\db(s-1)}\,.
\ee
Here symmetrization over the indices denoted by the same latter is implied and
to avoid projectors to the traceless and/or Young symmetry components we rose and lowered indices by means of $\gep^{\ga\gb}$, $\gep_{\ga\gb}$,
$\gep^{\da\db}$, $\gep_{\da\db}$.

\section{Conclusion}
Equations \eqref{spin_s_diff_eq} and gauge transformations \eqref{spin_s_gauge_trans} correspond to spin $s$ Fradkin-Tseytlin equations and gauge transformations
written in spinorial form. We, thus, shown that (independently of $\ga$) linear unfolded system constructed with the use of adjoint $\au(2,2)$-module
$\admodule_\ga$ and twisted-adjoint $\au(2,2)$-modules  $\twmodule_\ga$, $\btwmodule_\ga$ correspond to the collection of Fradkin-Tseytlin equations
for all spins $s=1,2,\ldots$ in one copy. Algebraically this is equivalent to decomposition formulas \eqref{addecompos}, \eqref{twdecompos}, which
show that independently of $\ga$ all adjoint modules $\admodule_\ga$ are isomorphic and the same is true for twisted-adjoint modules $\twmodule_\ga$.
We also found basis \eqref{iuga_basis}, \eqref{iuga_twbasis_real} for modules $\admodule_\ga$, $\twmodule_\ga$ in which this isomorphism become explicit.

However, the whole algebras $\aisu_\ga(2,2)$ are not isomorphic. This can be seen if one considers commutator of some its nonbilinear elements.
Thus, as follows from our analyses, the difference between these algebras become relevant on the nonlinear level only.

\end{document}